\documentclass{aa}
\usepackage{graphics}
\usepackage{psfig}

\begin{document}

   	\title{Numerical simulations of stellar SiO maser variability}

	\subtitle{Investigation of the effect of shocks}

 \author{E.M.L. Humphreys
          \inst{1}
           \and
          M.D. Gray
           \inst{2}
            \and
          J.A. Yates
           \inst{3}
            \and
          D. Field
           \inst{4}
            \and
          G. Bowen
           \inst{5}
            \and
           P.J. Diamond
            \inst{6}}

   \offprints{E.M.L. Humphreys, \email{liz@oso.chalmers.se}}

 \institute{ Onsala Space Observatory, 
               S-43 992 Onsala, Sweden\\
               email: liz@oso.chalmers.se      
         \and
               Department of Physics, 
               UMIST, PO Box 88, Manchester M60 1QD, UK\\
               email: mdg@saturn.phy.umist.ac.uk
	\and
	        Department of Physics and Astronomy,
 		UCL, Gower Street, London WC1E 6BT, UK\\
               email: jyates@star.ucl.ac.uk
         \and
               Institute for Storage Ring Facilities, University of Aarhus,
               DK-8000 Aarhus C, Denmark\\
               email: dfield@ifa.au.dk
 	\and
               Department of Physics and Astronomy, Iowa State University, 
               Ames IA 50011-3160, U.S.A.\\
               email: ghbowen@iastate.edu
	\and   
             University of Manchester, Nuffield Radio Astronomy Laboratories, 
             Jodrell Bank, Macclesfield, \\
		Cheshire SK11 9DL.  
             email: pdiamond@jb.man.ac.uk}

   \date{Received / Accepted }

   \abstract{
	A stellar hydrodynamic pulsation model has been combined with a SiO maser 
	model in an attempt to calculate  the temporal variability of SiO maser emission 
	in the circumstellar envelope (CE) of a model AGB star. This study
	investigates whether the variations in local
	physical conditions brought about by
	shocks are the predominant contributing factor to SiO maser
	variability because, in this work, the radiative part of the pump is constant. We find that
        some aspects of the variability are not consistent with a pump provided
        by shock-enhanced collisions alone.
        In these simulations, gas parcels of relatively 
	enhanced SiO abundance are distributed in a model CE by a Monte Carlo method, 
	at a single epoch of the stellar cycle. From this epoch on, Lagrangian motions 
	of individual 
	parcels are calculated according to the velocity fields encountered in 
	the model CE during the stellar pulsation cycle. The potentially masing gas parcels 
	therefore experience different densities and temperatures, and have varying 
	line-of-sight velocity gradients throughout the stellar cycle, which may or may not 
	be suitable to produce maser emission.
	At each epoch (separated by 16.6 days), emission lines from the parcels are combined 
	to produce synthetic spectra and VLBI-type images.
	We report here the results for $v=1$,  $J=1-0$ (43-GHz) and 
 	$J=2-1$ (86-GHz) masers and compare 
	synthetic  lineshapes and images with those 
	observed.  
	Strong SiO maser emission is calculated to form in an 
	unfilled ring within a few stellar radii of the photosphere, 
	indicating a tangential amplification process. 
	The diameter of the synthetic  maser ring is dependent upon stellar phase, as clearly
	observed for TX Cam, and upon maser transition. Proper motions of brightly masing parcels 
	are comparable to measurements for some maser components in R Aqr and TX Cam, although we
	are unable to reproduce all of the observed motions.  
	Synthetic lineshapes
	peak at the stellar velocity, have typical Mira linewidths and vary in intensity with 
	stellar phase. 
	However, the model fails quantitatively in several respects.
	We attribute these failings to (i) lack of an accurate,
        time-varying stellar IR field
        (ii) post-shock kinetic temperatures which are too high, due to the cooling function
	included in our model and (iii) the lack of a detailed treatment of the
	chemistry of the inner CE. We expect the use of oxygen-rich hydrodynamical
 	stellar models which are currently under development to
	alleviate these problems.
}

	\titlerunning{Numerical simulations of stellar SiO  masers}

      \maketitle


\section{Introduction}
\label{s:intro}

	It is well-known from VLBI experiments that bright $v=1$ $J=1-0$ SiO 
	maser components in M-Mira and Supergiant stars lie in approximate ring-type 
	structures of several AU in diameter (in the range 1.5 -- 4.0 R$_{*}$), 
	which are assumed to be centered on the stellar position 
	(Colomer et al. \cite{colomer92};  Diamond et al. \cite{diamond94}; 
	Greenhill et al.\cite{greenhill95}; Miyoshi et al. \cite{miyoshi}; 
	Boboltz et al. \cite{boboltz}; Doeleman et al. \cite{doele} at 86-GHz; 
	Desmurs et al. \cite{desmurs}; Yi et al. \cite{yi}). The width of the 
        projected ring of clumpy emission 
	is relatively narrow. The ring diameter changes with 
	stellar pulsation phase, 
	as emission brightens and dims,  during the cycles of M-Mira stars such as 
	TX Cam and R Aqr (Diamond \& Kemball \cite{diamond99}; Boboltz et al. \cite{boboltz}). 
	M-Miras and other classes of star on the thermally pulsing AGB (TP-AGB) are of 
	interest due to their high mass loss rates (in the range 10$^{-7}$ -- 10$^{-4}$ 
	M$_{\odot}$yr$^{-1}$) which replenish the interstellar medium with heavy elements 
	and dust grains. SiO masers, which may  now be studied at 43-GHz with an angular 
	resolution of 200$\mu$as and a velocity resolution of 0.1 kms$^{-1}$ using the 
	VLBA, provide excellent probes of the dynamics and time-varying physical conditions 
	of the extended atmosphere/inner CE of AGB stars. A detailed VLBA study of $v=1$ 
	$J=1-0$ (43-GHz) masers in the Mira TX Cam provides data on the inner CE at 50 epochs 
	over 1.25 stellar cycles (Diamond \& Kemball \cite{diamond99}).

	An important role in the mass loss mechanism of
	TP-AGB stars is thought to be played by a combination of stellar 
	pulsation-driven shock waves and 
	radiation pressure on dust. In this scenario, a succession of shocks 
	greatly extends the Mira atmosphere, and material is accelerated, overcoming 
	the gravity of the star with accompanying mass loss. Observational evidence 
	for shock waves is provided by large amplitudes ($\Delta v$ = 20 -- 30 kms$^{-1}$) 
	in the velocity curves derived from photospheric CO lines. These 
	indicate that an outwardly propagating shock wave is associated 
	with the stellar pulsation (Hinkle et al. \cite{hinkle84}; 
	\cite{hinkle97}). 
	However models are unable to reproduce the relatively high
	mass loss rates measured for TP-AGB stars without the inclusion of 
	radiation pressure on dust, which must be formed close to the star
	(Bowen \cite{bowen88}). 
	This is possible in the shocked model in which relatively dense
	gas is levitated to regions of lower temperature by the shocks, 
	allowing dust condensation to take place within a few stellar radii 
	from the photosphere.
	Radiation pressure from stellar photons then accelerates the grains, 
	driving a slow, cool wind through frictional coupling with 
	the circumstellar gas.
	Infrared interferometry measurements support the formation of dust 
	shells with inner radii of 3 -- 5 R$_{*}$ towards a number of Miras
	and Supergiants (Danchi et al. \cite{danchi}).
	Towards the Supergiant VX Sgr, coordinated SiO maser and infrared 
	interferometry observations indicate that the $v=1$ $J=1-0$ (43-GHz) ring of 
	bright SiO masers and the inner radius of the dust shell are 
	well-separated, the masers lying  at $\sim$3 R$_{*}$ inside the 
	dust region of the supergiant star (Greenhill et al. \cite{greenhill95}).

	Much remains to be understood about the complex processes in these stars.
	For example, recent multi-wavelength observations of the S-type Mira, $\chi$ Cygni, show 
	that the star is largest at minimum light, whereas current non-linear pulsation models 
	predict the maximum physical size to  occur close to maximum light 
	(Young et al. \cite{young}). With respect to the inner CE region, measurements 
	of the 8-GHz ``radio photospheres''
	of several stars indicate that shocks must have been 
	significantly damped by the inner radius of the SiO maser zone 
	and propagate outwards with velocities of $<$5 kms$^{-1}$
	(Reid \& Menten \cite{reid}).
	Yet proper motion measurements for the SiO masers in TX Cam show that some
	components are outflowing at velocities of $\sim$10 kms$^{-1}$ (Diamond, 
	private communication). In addition, magnetic fields in the SiO maser zone may be
	of the order of several gauss (Kemball \& Diamond \cite{kemball}), and will 
	therefore play a crucial role in the gas dynamics of the inner CE, or may be 
	only a few tens of milligauss
	(Wiebe \& Watson \cite{wiebe}). Finally, recent observations appear to have detected
	rotation in the SiO maser zone of some stars (e.g. Boboltz \& Marvel \cite{boboltz00}).

	As a first essay towards understanding some of these observations, 
	we combine a model of a pulsating TP-AGB star, in which
	mass loss is driven by shocks and radiation pressure on dust, with an SiO 
	maser model. Humphreys et al. (\cite{humphreys}; hereafter H96) showed how, 
	at a single stellar phase, the key features of SiO maser emission are 
	reproduced by such a model. In the present study, the crude 
representation of the stellar IR radiation field results in a decoupling of
the effects on SiO masers of a varying IR radiation field and
	shocks in the inner CE.
        The aim of the present study is to perform
	simulations of the time evolution of SiO masers, both in intensity
	and in spatial distibution, throughout a stellar pulsation cycle.
        Although constant radiative pumping is present, we effectively
	determine whether the effect of shocks alone is sufficient to
	reproduce the observed features of SiO maser variability.


\section{SiO maser temporal variability}
\label{s:observations}

	Three types of variability may be identified. Firstly, there is variability 
	from cycle to cycle, random in behaviour in the sense that SiO maser spectra 
	are quite different in appearance from one cycle to the next. Secondly, there 
	may be relatively orderly long-term variability within a cycle, in which maser 
	flux passes through a maximum at an optical phase not far removed from maximum 
	light, as described in Sect.~\ref{ss:longterm}. Thirdly there is rapid variability 
	over a period of a day to a few tens of days, as described in Sect.~\ref{ss:rapid}. 
	We are concerned mainly in the present paper with the second and third types of 
	variability, which we refer to as long term and short term (or rapid) variability. 
	We do not address variability from cycle to cycle in this work, except for an 
	investigation of the stellar phase at which maser emission is severely
	disrupted (Sect.~\ref{r:images}).
	
	The long term variability of SiO masers has been investigated in monitoring 
	programs towards many Mira variables, for example $o$ Ceti (Mira), R Aqr, 
	TX Cam, U Her, R LMi, IK Tau (=NML Tau), R Leo, W Hya, $\chi$ Cygni, 
	R Cas, U Ori, R Aql, R Cnc, X Hya and T Cep. The most complete sets of observations 
	performed to date are those of Mart\'{\i}nez et al. (\cite{martinez88}; hereafter MBA88)
	and Alcolea et al. (\cite{alcolea99}) for $v=1$ 43-GHz masers 
	and 
	Nyman \& Olofsson (\cite{nyman}; NO86 hereafter) for $v=1$ 86-GHz masers.
	These studies demonstrate long and  short term variability. A 
	statistical approach to the study of long term variability has been 
	taken by papers involving Cho. The same maser transition was observed 
	towards a large sample of stars, yielding statistical information on 
	that maser as a function of stellar phase. Very short term variability 
	on a timescale of 12 to 24 hours measured over a small part of a stellar 
	cycle has been investigated in one detailed study of R Cas and R Leo 
	(Pijpers et al. \cite{pijpers94}) at 43-GHz $v=1$ $J=1 - 0$.  
	We consider first the observational data for long term variability. 

\subsection{Long term variability: correlations with optical phase of the host star}
\label{ss:longterm}

The following general characteristics can be derived from observations of Miras.

\medskip

\noindent (i) The period is stable from cycle to cycle, but spectra 
	vary greatly (NO86). Indeed, so do 
	photospheric sizes, as measured by speckle interferometry 
	(Bonneau et al. \cite{bonneau}).

\medskip

\noindent (ii) Optical, infrared and low frequency (43 and 86-GHz) SiO maser fluxes are
	correlated. There is firm evidence for an average phase lag of $\sim$0.2 of a period 
	between optical and maser maxima, but this may vary between objects, and for the same 
	object, between cycles (MBA88, Cho et al. \cite{cho96b}; Alcolea et al. \cite{alcolea99}). 
	From the 43-GHz data in MBA88 the phase lag had a mean of 0.18, standard deviation 0.1. 

\medskip

\noindent (iii) SiO maser emission appears to vary in phase
	with near and mid-IR lightcurves. Alcolea et al. (\cite{alcolea99})  
	show that the SiO maser maxima are closely 
	related to the light curve at 3.79 and 4.64 $\mu$m for R Aqr and IK Tau. However, 
	the amplitudes of the IR and maser lightcurves are not correlated.
A good correlation of the maxima of 86\,GHz masers and the IR continuum at
1.04 $\mu$m was found in 6 out of 8 miras studied in NO86, but failed for two
objects, R Leo and R Aql. 

\medskip

\noindent (iv) SiO masers often have strong linear polarization 
	(Barvainis \& Predmore \cite{barvainis}) which may introduce bias 
	into observations made with a single linearly polarized feed. 
	NO86 and MBA88 argue the bias is small however. Papers involving Clark
	are the best in this respect, since they measure the Stokes parameters I, Q and U.

\medskip

\noindent (v) The contrast between the highest and lowest peak lineshape intensity 
	(and lineshape area) during a cycle is very variable from cycle to cycle and 
	star to star. 
        For the Mira-type stars studied at 43\,GHz in 
	Alcolea et al. (\cite{alcolea99}), this value varied between 1 up to as high as 20, 
	in the
	case of $o$ Ceti.
However, at 86\,GHz the dynamic range during one cycle of $o$ Ceti was at
least 100, and could have been considerably larger, since masers in this source
fell below the 3$\sigma$ detection limit of 5\,Jy near minimum.

\medskip

\noindent (vi) The velocity extent of emission is typically $\sim$15 kms$^{-1}$, with 
	the cycle-averaged peaks of the spectra close to the stellar velocity (NO86). 
	Cho et al.(\cite{cho96b}) find the mean velocity of their sample was 
	red-shifted with respect to V$_{*}$ for optical phase 0.3 to 0.8 with blue-shifted 
	emission appearing from $\sim$0.85 and dominating between 0.0 and 0.2. The cycle-averaged 
	spectral peak at $v=1$, $J=1-0$ was 0.3 kms$^{-1}$ to the red of V$_{*}$, compared 
	to the blueshift of  $\sim$1 kms$^{-1}$ found by NO86.
	In addition, SiO profiles sometimes have unusually broad, low-intensity
	linewings which can exceed the expansion velocity of the circumstellar
	envelope as traced by thermal CO measurements, at least at 86-GHz
	(Herpin et al.\cite{herpin98}).

\subsection{Short term variability}
\label{ss:rapid}
        
	Balister et al.(\cite{balister}) noted that a number of SiO maser sources 
	(in $v=1$, $J=1 - 0$  at 43-GHz) showed variability over a period of a few 
	days. Spectra are shown of the semiregular variable R Dor and the supergiant 
	AH Sco. Both stars showed a marked increase in maser flux, especially AH Sco, 
	over a period of only 4 days. This remarkable phenomenon was not however studied 
	in any detail until the work of Pijpers et al. (\cite{pijpers94}) who observed R Cas 
	and R Leo over phases 0.96 to 0.09 and 0.14 to 0.23 respectively. With respect to 
	total flux, R Cas showed a slow decline and R Leo went through a marked maximum over 
	this period of 40 days. In both cases these variations probably represented correlation 
	with the stellar phase, as discussed in Sect.~\ref{ss:longterm}, rather than short term 
	variability. Line profiles however changed within a period of days. Channel by channel 
	variations ranged between 10\% and 30\% of the peak flux, with typical timescales of 
	variation of 10 to 20 days. The velocity centroid of the dominant maser peaks apparently 
	shifted by $\sim$1 kms$^{-1}$ on the same timescale. No clear periodicity was observed 
	in these short term variations.

\subsection{Survival of masers during a stellar cycle}
\label{ss:survival}

	Clark et al. (\cite{clark84a}) and Clark et al. (\cite{clark85}) decomposed their 
	86-GHz maser spectra into a number of overlapping gaussian components. 
	They find for R Leo and W Hya that individual gaussian components in the 
	Stokes I spectra often persist in successive spectra over the greater part 
	of an optical cycle. Miller et al. (\cite{miller84}) reach the same conclusion 
	for $o$-Ceti. The consistency of polarization position angle in these individual 
	components is an important factor in reaching this conclusion. Given that the centre 
	velocity, velocity width, polarization position angle and fractional polarization of 
	individual features persist throughout much of the optical cycle, Clark et al. 
	(\cite{clark85}) argue that individual maser zones are stable and discrete physical 
	entities. The continuity of maser features was reported to be broken at, or just 
	prior to, maximum light for W Hya (Clark et al. \cite{clark85}), for R Cas (Clark et 
	al. \cite{clark82a}), for R Leo (Clark et al. \cite{clark82b}; 
	Clark et al. \cite{clark84a}) and for $o$-Ceti (Miller et al. \cite{miller84}). 
	NO86 however found no such result for R Cas, R Leo and $o$-Ceti, but note that a 
	lack of polarization information (and line blending) could have caused the effect 
	to be missed. According to MBA88 and Alcolea et al. (\cite{alcolea99}) however, 
	continuity appears to be broken where SiO maser brightness is at minimum, at around 
	an optical phase of 0.7, rather than near optical maximum. 

\subsection{Proper motions during a stellar cycle}
\label{ss:VLBI}

	VLBI studies have been made towards several M-Miras: 
	R Aqr (Boboltz et al. \cite{boboltz}; Hollis et al. \cite{hollis} for $v=1$ 43-GHz masers); TX Cam 
        (Diamond \& Kemball \cite{diamond99} for $v=1$ 43-GHz masers;
	Desmurs et al. \cite{desmurs} for $v=1$ and $v=2$ 43-GHz masers;
	Yi et al. \cite{yi} for $v=1$ and $v=2$ 43-GHz masers); $o$ Ceti 
	(Gardner et al. \cite{gardner} for $v=1$ and $v=2$ 43-GHz masers); R Cas 
	(Yi et al. \cite{yi}  for $v=1$ and $v=2$ 43-GHz masers). 
 	Boboltz et al.(\cite{boboltz}) made the first proper motion measurements of 
	circumstellar SiO masers towards R Aqr. They observed an infalling mean component 
	proper motion of 1 mas in 98 days, corresponding to 
	0.22 AU over 0.25 stellar periods or 4 kms$^{-1}$,  for their assumed 
	distance to R Aqr of 220 pc.
 	In the long-term monitoring of TX Cam, more
	complex motions are evident.
	Maser components are generally expanding 
	outwards from the star at a typical velocity of 3.65 kms$^{-1}$ 
	(Diamond \& Kemball (\cite{diamond99}), although some have  constant 
	outflow velocities as high as  $\sim$10 kms$^{-1}$. In one 
	quadrant of the images, components decelerate smoothly as they appear 
	to run into denser material. Some of the ring also shows features sliding in
	non-radial directions. Masers in one region of the ring  
 	fall inwards towards the star and then appear to ``bounce back'', 
	an effect which could be due to  a new shockwave moving through the maser zone
	(Diamond, private communication). 
	Disruption to the maser ring appears to occur at 
 	maser minimum light at around optical phase 0.67 (Diamond, private communication).
	A new maser emission ring, of smaller angular extent than the disrupted
	ring, is then observed to appear.

\begin{figure}
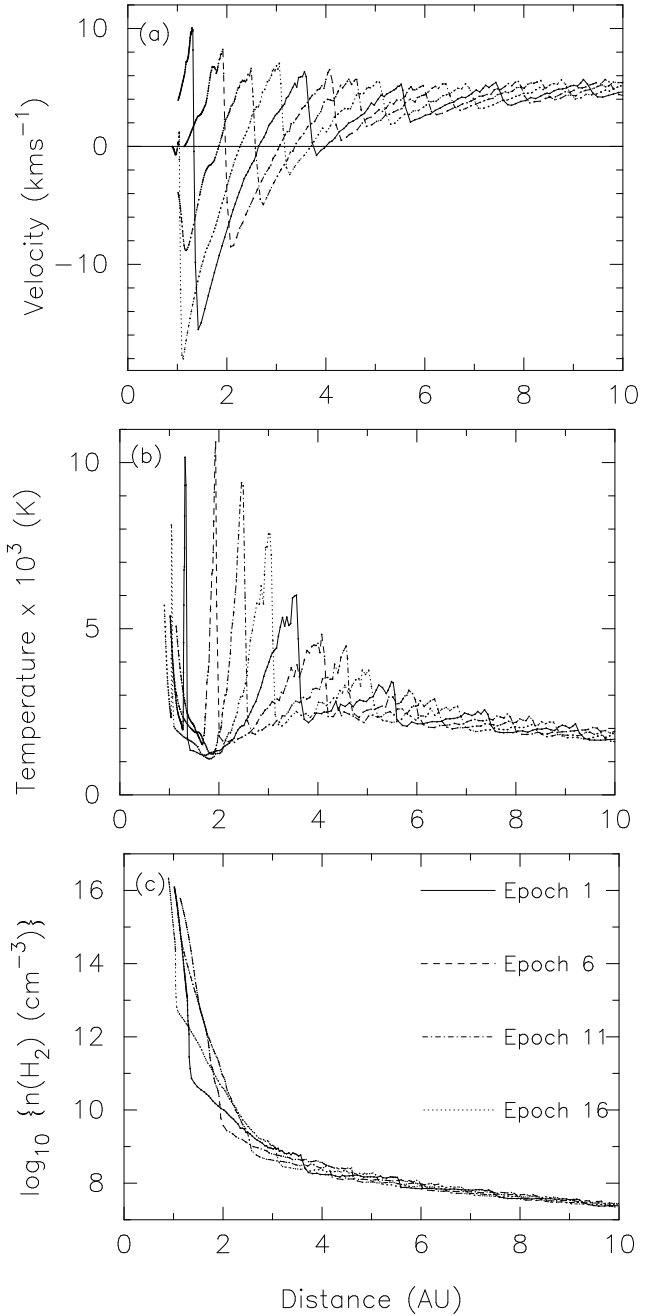

\begin{center}
\vspace{0.1cm} \hspace{0.1cm} \psfig{file=siovarvel.ps,width=83mm,angle=270}
\vspace{0.3cm} \hspace{0.1cm} \psfig{file=siovartem.ps,width=83mm,angle=270}
\vspace{0.3cm} \hspace{0.1cm} \psfig{file=siovarden.ps,width=83mm,angle=270}
\end{center}
\vspace{-0.5cm}
\caption{\label{f:bowenplot} 
	(a) The variation with distance of the radial velocity of material in the circumsteller  	envelope (CE) of a M-Mira variable for 4 model epochs, obtained using the 
	hydrodynamic-pulsation model of Bowen described in Sect.~\ref{ss:bowenmodel}. 
	The model epochs shown are 1, 6, 11 and 16  corresponding 
 	respectively to 0, 83, 166 and 249 days in the stellar pulsation cycle. 
	The characteristics of the model 
 	M-Mira are given in Table~\ref{t:mirainputs}. (b) As for (a), but showing 
the variation of gas kinetic 
 	temperature. (c) As for (a), 
	but showing the variation of gas number density.
}
\end{figure}

\section{The model of SiO masers in the circumstellar environment}
\label{s:siomodel}

	The technical means employed in this paper have been described in 
	some detail in H96. A hydrodynamic pulsation model for M-type Miras, based 
	upon that in Bowen (\cite{bowen88}, \cite{bowen89}), is coupled to a model 
	for SiO maser action based on that described in Doel et al. (\cite{doel95}; 
	D95 hereafter) and Gray et al. (1995). H96 considered only a single phase of the stellar pulsation 
	and showed how, at this phase, the physical conditions in the CE were such 
	that a ring of SiO masers should form about the star at a distance from the 
	photosphere of $\sim$1 R$_{*}$ (1 R$_{*}$ = 1.1 AU). The synthetic map produced 
	showed an encouraging resemblance to data acquired using the VLBA 
	(Diamond et al. \cite{diamond94} and new data presented in H96). 
	The lineshapes of masers in various transitions in v=1 and v=2 ranging 
	from $J=1 - 0$ to $J=7 - 6$ were also presented and were found in form and 
	width to resemble lines observed in the very large number of sources in the 
	literature. H96 allowed certain conclusions to be drawn with respect to the 
	location of SiO masers, that is, that they do not in general reside in the 
	stellar wind, defining the stellar wind as that part of the flow in which 
	the radial velocity of the CE is always directed away from the photosphere 
	with no further episode of infall. The agreement between the calculated and 
	observed position of SiO masers lends support to the shock driven pulsation 
	model, since it is this model which determines spatial variation of number 
	density, temperature and velocity field which in turn dictates where the SiO 
	masers should be found.

	The present work is an extension of H96 involving calculations at 20 phases 
	of stellar pulsation, carrying the evolution of the CE through a complete 
	stellar cycle. As the cycle develops, the number densities, kinetic temperatures 
	and bulk velocity fields change as a shock propagates through the photosphere 
	and into the CE. This in turn causes populations of rovibrational levels of SiO 
	to be modified. Inversions between rotational levels therefore 
	increase, decrease or disappear and maser emission is modified accordingly. The 
	remainder of this section is devoted to a description of the hydrodynamic pulsation 
	and the kinetic and radiative SiO maser models.

\begin{table}

\begin{center}
\vspace{0.5cm}
 \begin{tabular}
 {ll}
\hline
                               &      \\
\multicolumn{2}{c}{Parameters of the model M-Mira}\\
                               &      \\
 \hline
\\
Mass                      &          1 M$_{\odot}$          \\
Fundamental Period                        &  332 days           \\
Stellar Radius                        &   244 R$_{\odot }$ (1.7 x 10$^{11}$ m )  \\
Effective Temperature                        &   3002.2 K           \\
Maximum Inner Boundary Speed	          &    3.93 kms$^{-1}$     \\
Mass Loss Rate & 1.8 x 10$^{-7}$ M$_{\odot }$yr$^{-1}$\\
\\
\hline
\end{tabular}
\end{center}
\caption{\label{t:mirainputs}
Characteristics of the model star used to compute the physical conditions in the CE of 
a M-Mira long-period variable.
}
\end{table}

\subsection{The hydrodynamic pulsation model}
\label{ss:bowenmodel}

	The model of the time-varying CE is based upon that described in 
	Bowen (\cite{bowen88},\cite{bowen89}) and is identical to that 
	described in H96. The models of Bowen have been chosen over more recent 
	models, which include time-dependent dust formation, such as the work of 
	Bessell, H\"{o}fner and Fleischer, as they have tended to concentrate on 
	carbon-rich variables (Fleischer et al. \cite{fleischer91}, \cite{fleischer92}; 
	Bessell et al.\cite{bessell}; H\"{o}fner et al.\cite{hofner96}; H\"{o}fner \cite{hofner99}), 
	which clearly do not  support SiO masers since the greater part of the 
	oxygen budget is bound up in CO. The models by Bowen represent oxygen-rich 
	Miras, which comprise the vast majority of the AGB star population, although 
	dust in these models is introduced as a parametrised opacity.

	Briefly, the stellar oscillation is driven by a sinusoidally varying force situated 
	below the photosphere of the star. Radiation pressure is assumed to act on 
	dust, and on H$_{2}$O molecules. The quantity of dust present depends on the 
	grain temperature, with no dust for T$_{rad}$ = 2000 K and a maximum 
	achieved for T$_{rad}$ $<$ 1000 K. The outflow of material from the star becomes 
	a steady wind at 40 to 50 R$_{*}$, the outer boundary of the model. The same stellar 
	parameters are used as in H96 and for ease of reference these are shown again 
	in Table~\ref{t:mirainputs}. The stellar phase in the model (the `model phase') is 
	defined as zero when the sinusoidally varying inner boundary is moving outwards with 
	maximum velocity.  We refer to the 
	problem of relating the model phase to stellar optical phase in 
	Sect.~\ref{s:results}.

\begin{figure}[b]
\psfig{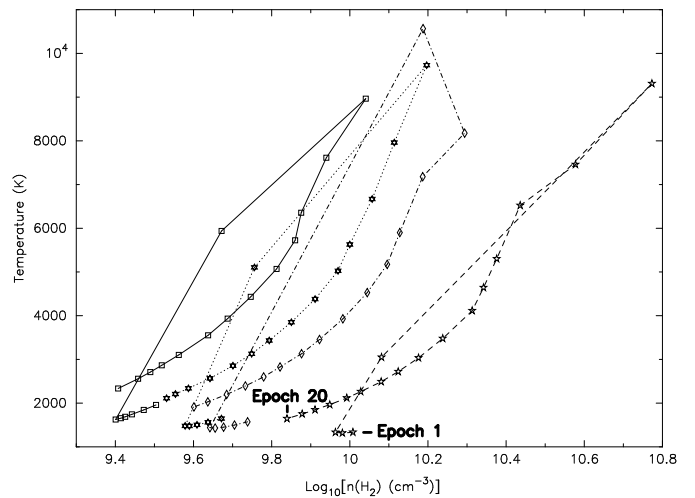}
\caption{\label{f:phasediag}
The number density - temperature space which 4 masing parcels of gas 
experience during a stellar cycle. 
Points are separated by an interval of 16.6 days. 
The start and end of the cycle is marked for one of the parcels. 
}
\end{figure}

	In the CE, a parcel of gas for example close to the photosphere, initially experiences 
	a brief but very strong outward acceleration, and thereafter decelerates, reversing 
	direction. After a period of infall it is then driven outwards again and this cycle 
	is repeated with successively weaker impulses as the material finds itself increasingly 
	further from the photosphere. Thus if one were arbitrarily to choose some initial 
	radial position 
	within the CE, and ride on that parcel of gas in the co-moving Lagrangian frame, then 
	one would experience a continuously changing set of number densities, temperatures and 
	velocity fields. The nature of these changes would depend  on the initial choice of 
	radius. This is illustrated in Fig.~\ref{f:bowenplot}\,a,b and c and 
	in Fig.~\ref{f:phasediag}. 
	Fig.~\ref{f:bowenplot}a shows the variation in velocity at four different model phases, 
	Fig.~\ref{f:bowenplot}b the variation in temperature and Fig.~\ref{f:bowenplot}c	   	   
	the variation in number density. The model phases chosen correspond to zero, 
	83 days, 166 days 
	and 249 days, the stellar pulsational period being 332.0 days. Values are plotted out to 
	10 AU only, in order to show the zone of interest for SiO masers.
	Any vertical line drawn through the four graphs in each figure illustrates how the 
	chosen parameter would vary with stellar phase. 
	Fig.~\ref{f:phasediag} 
	illustrates the density and temperature history, in a Lagrangian frame,    
	for four masing parcels of gas during a stellar cycle i.e. 
	Fig.~\ref{f:bowenplot} gives an Eulerian view, and Fig.~\ref{f:phasediag} the 
	Lagrangian view following a particle. 
	Points in 
	Fig.~\ref{f:phasediag} are separated by 16.6 days. Similar phase plots may be 
	made for velocity-temperature and velocity-density.


\subsection{Overview of the computational procedure}
\label{ss:compoverview}

	The calculation of maser spectra and of maps proceeds as follows. Sites of maser 
	action in the CE are first chosen on a random basis, for one epoch
	of the calculations only. 
	At present, it is assumed for simplicity that maser zones 
	maintain their integrity throughout the cycle involving model phase zero to 1. 
	The trajectories of these initially chosen sites are followed in terms of their 
	radial distances from the star as the cycle progresses, using data from the stellar 
	pulsation model. At any chosen phase, for each radial distance a number density, 
	kinetic temperature and velocity field may be assigned, as is apparent from 
        Fig.~\ref{f:bowenplot}a,b,c and Fig.~\ref{f:phasediag}. For each such distance, 
	SiO rovibrational level populations are calculated. If these populations lead to 
	inversions, the intensities of the resulting masers are calculated for paths in 
	our line-of-sight. In this way, maps of SiO maser emission may be calculated for 
	as many phases of the stellar cycle as desired and for any masing transition. The 
	series of steps involved in synthesising maser images is described in more detail in 
	the succeeding sections. Maser spectra are generated at any phase by summing over the 
	contributions of  individual maser spots, with line-of-sight velocity shifts appropriate 
	for their position.
 
\subsection{Choice of masing zones: proper motions in the stellar cycle}

	VLBI images of SiO emission at 43-GHz and 86-GHz clearly show that at some chosen 
	radial distance from the photosphere only certain portions of gas in the CE are 
	capable of sustaining maser action. This suggests that there exists small-scale, 
	non-radial velocity structure superimposed on the large-scale velocity structure 
	calculated by the stellar pulsation model and/or inhomogeneity in the density and
	temperature structure of the CE, such as clump formation via thermal instabilities 
	(Cuntz \& Muchmore \cite{cuntz}). Since the nature of such small-scale 
	structure is unknown, sites of maser action were chosen randomly in H96. This 
	turned out to be a successful means to reproduce the appearance of VLBI maps 
	(of TX Cam) and we adopt the same procedure here. Thus, as in H96, we choose sites 
	of maser action in the CE at random radial distances from the star within a 
	specified range of radius, $r$, and in fact we start with the same set as in H96. 
	Only directions of maser propagation in our line-of-sight are considered 
	(see Sect.~\ref{ss:maserprop}). The range of radius over which values are 
	randomly chosen extends from an inner boundary r$_{in}$ = R$_{*}$ = 1.7x10$^{11}$ m 
	to an outer boundary r$_{out}$ = 5 R$_{*}$. Tests with larger values of r$_{out}$ 
	showed that this range of radius was sufficient to include all masers in all transitions. 
	The number of random values of $r$ used is a free parameter of the model. 
	We found in H96 that 
	1500 values, of which $\sim$25$\%$ yielded inversions, was appropriate to reproduce the 
	appearance of VLBI maps of TX Cam in terms of the number of bright maser spots around the 
	host star. The same value was chosen in the present study. Maser propagation is discussed 
	in more detail in Sect.~\ref{ss:maserprop}.
        
	The CE is seen in projection in the plane of the sky and the location of masing zones 
	within the spherical CE must be fully specified in order to reproduce VLBI maps. We use 
	spherical polar coordinates r, $\theta$ and $\phi$, choosing random points in such a way 
	as to provide uniform sampling by volume. Thus a set of random r, $\theta$ and $\phi$ is 
	chosen and weighted as described in detail in H96, giving r', $\theta$' and $\phi$',  
	where 
	primed quantities refer to weighted values. In summary, r' defines the physical conditions 
	and $\theta$' and $\phi$' are used to locate the position, in the plane of the sky, of any 
	maser spot that may develop. The next stage is to advance the value of radius r' to that 
	corresponding to the next phase for which the physical conditions and maser location are 
	required. The distance $\delta$r' that any mass of gas of velocity V travels in time $t$ is given by

\begin{equation}\rm
\delta r' = \int^{t_{0} + \,\delta t}_{t_{0}} dt \,V\,( r' , t )
\,,
\end{equation}

\noindent which is computed using $ \delta r' = \delta t \,V ( t = t_{0} ) 
 	+ \delta t^{2}/2 \,dV/dt + \delta t^{3} / 3! \,d^{2}V/dt^{2} $ etc
	where the differentials are evaluated at $t = t_{0}$. The total 
	differentials $dV/dt$ and $d^{2}V/dt^{2}$ are obtained from the output of 
	the stellar model as follows. The pulsation model describes how the velocity 
	at a given r' changes as phase advances and how for any one phase the velocity 
	changes with radial distance in the neighbourhood of the radial position r'. These 
	data may then be used to obtain $dV/dr'$ and $d^{2}V/dr'^{2}$ through an expansion in 
	partial differentials, where differentials up to third order have been included. 
	Variations of r' with phase trace out the proper motions of potential maser sites 
	in the course of the stellar cycle.
        The CE model generates 520 values of number density, temperature and velocity 
	over a range of radial distance 1.5 x 10$^{11}$ m to 7.3 x 10$^{12}$ m from the 
	centre of the star. The physical conditions associated with any position are 
	obtained through linear interpolation. Conditions are generated for each of the 
	1500 values of r', $\theta$' and $\phi$' for each of the 20 phases studied, making 
	in all 30,000 sets of conditions. A certain proportion of the 1500 sets of conditions 
	at any phase yield population inversions, and excluding the very small proportion which 
	lie behind the disk of the host star, one may trace the proper motions of brighter persistent 
	masers in the CE, as described in Sect.~\ref{r:images}. In this connection, SiO 
	rovibrational populations 
	assume values corresponding to the prevalent physical conditions on a timescale which is very 
	rapid compared with the timescale of variation of the physical conditions. Collisional and 
	radiative events determine the SiO populations, the rate-determining steps being collisional. 
	A typical rate coefficient for the transfer of rovibrational energy through collisions of 
	H$_{2}$ with SiO may be of the order of 10$^{-12}$ cm$^{3}$s$^{-1}$ or greater. The number 
	density lies between 10$^{8}$ and 10$^{10}$ cm$^{-3}$ in a typical maser zone, and thus the 
	timescale for pumping maser inversions does not exceed a few x 10$^{4}$ seconds and is 
	typically rather less than 10$^{4}$ seconds. Thus the SiO populations effectively respond 
	instantaneously to changes in the environment.

\subsection{Calculation of rovibrational level populations}
\label{ss:levpops}

	The methods used for the calculation of the populations of SiO rovibrational levels 
	have been described in detail in D95 with modifications described in H96. 
	Populations of levels are initially calculated 
	ignoring the presence of inversion in the medium or the influence of maser radiation.  
	In this connection, population inversions involve rotational transitions whose line 
	emissivities are negligible by comparison with those of rovibrational transitions 
	(H96; Yates et al. \cite{yates97}). The effects of the presence of masers are discussed 
	in Sect.~\ref{ss:maserprop}. A discussion of the work of other groups involved in 
	modelling 
	SiO maser emission (e.g. Langer \& Watson \cite{langer}; Lockett \& Elitzur \cite{lockett}; Bujarrabal \cite{buj94a}, 
	\cite{buj94b}) is given in D95.
        Calculation of the SiO populations involves solution of the master equations for 
	200 rovibrational energy levels of SiO, involving $v$ = 0 -- 4 and $J$ = 0 -- 39 in each 
	vibrational state. 
Stellar continuum radiation is included with a geometrical dilution 
	factor of 0.1 (D95). As the model has a fixed photospheric temperature,
with no phase variation (see Table~1.), and is represented as a black-body,
the radiative part of the pumping scheme is represented only crudely. We
note that the present hydrodynamic model provides no means of tracing either
the true radius or temperature of the photosphere in the near to mid-IR
wavelength range. Consequently, the present combined model is not capable
of testing the observed phase correlation between the IR continuum and 
low-frequency SiO maser peaks. The current work is therefore a useful
test, in that it provides information about those aspects of maser
time variability which can and cannot be reproduced by a pump in which 
collisions provide that part of the pumping scheme which is dependent on
stellar phase. 
 An infrared dust radiation field is also present. The dust continuum 
	is represented in a crude manner by a black-body function at a single dust temperature of 
	200 K modified by a wavelength dependent factor, 
	$(\lambda/\lambda_{0})^{-p}$, where $\lambda_{0}$ is 80$\mu$m
	and $p = 1.1$ (Rowan-Robinson et al.\,1986), and the shortest value of $\lambda$ is 
	$\sim$ 8 $\mu$m. D95 shows that the only significant property of the dust radiation 
	field (or any other external field) is that it should be weak. Thus our very simple 
	and inaccurate representation of the dust field is not a significant source of error 
	in computation of the SiO energy level populations except for high $v$-states and minor 
	isotopomers. The dust radiation field is spatially diluted through a factor of 0.01, since 
	it is assumed that significant quantities of dust are not found in regions occupied by SiO 
	masers (Greenhill et al. \cite{greenhill95}). The dust may also be patchy, with a covering 
	factor significantly $<$1 (D95). Collisional rate coefficients for the transfer 
	of rotational 
	and rovibrational energy between SiO and its collision partners (see below) have been 
	taken 
	from Bieniek \& Green (\cite{bg83a},\cite{bg83b}; hereafter BG83) where values have been 
	extrapolated, as 
	described in D95 and Doel (\cite{doel90}), to include all $v$ and $J$ states in 
	the maser model. We note that Lockett \& Elitzur (\cite{lockett}) also employed the 
	rate data of BG83, and 
        extended the range of vibrational transitions treated up to $\Delta$v = 4.
        Langer \& Watson (\cite{langer}) used the BG83 data for collisions
        of molecular hydrogen with SiO,  and
          additionally estimated rates for collisions of SiO with
       atomic hydrogen.
 The Sobolev or large velocity gradient (LVG) approximation is used to 
	calculate self-consistent 
	populations and line and continuum radiation fields, as in 
	Lockett \& Elitzur (\cite{lockett}). 
	In support of the use of this approximation, CE models clearly indicate the presence of 
	supersonic velocity shifts over distances smaller than or comparable to the dimensions of 
	the maser zone. However the physical conditions in the CE close to the photosphere change 
	markedly over a typical Sobolev length, whilst in the simple form of LVG used here 
	conditions 
	are assumed constant over a Sobolev length (= Doppler width divided by the local velocity 
	gradient). The LVG approximation evidently introduces a coarse-grained interpretation of 
	observational data. As described in H96, the number density may change by as much 
	as an order 
	of magnitude over the Sobolev length, representing a rather severe spatial 
	averaging. As in H96 
	we use a general expression for the angle-averaged photon escape probability 
	in a spherically 
	symmetric system, using values of the radial velocity gradient $dv/dr'$ and the tangential 
	velocity gradient $v/r'$ taken from the stellar pulsation model for each value of radius. 
	As we discuss in H96 we ignore the influence of non-monotonic velocity gradients, which 
	may cause a ray of light to encounter additional surfaces in the medium at equal velocity 
	to an emitting point (Rybicki \& Hummer \cite{rybicki}). D95 concluded that masers are 
	pumped by  collisions at typically 1500 K and a number 
	density of $\sim$5 x 10$^{9}$ cm$^{-3}$, 
	in an environment containing large velocity gradients, perhaps in excess of 
	10$^{5}$ kms$^{-1}$pc$^{-1}$, and experiencing only a weak dust continuum radiation 
	field. D95 pointed out that these conditions were consistent with those predicted by 
	CE models (e.g. Willson \cite{willson}; Bowen \cite{bowen88}; Bowen \cite{bowen89}; 
	Bowen \& Willson \cite{bowen91}; Fleischer et al. \cite{fleischer92}) for a zone lying 
	within about a stellar radius from the photosphere.

        The chief sources of error in the calculation of the populations of SiO energy levels are:

\noindent (i) the use of inaccurate rate coefficients for energy transferring collisions 
	between SiO and collision partners. Calculated values of BG83
	are for He. Collision partners in the CE are 
	H$_{2}$ and H, in unknown proportions, with a relatively small contribution 
	from He. 

\noindent (ii) use of the LVG approximation. Exact methods, omitting velocity gradients, 
	have been used in Bujarrabal (\cite{buj94a},\cite{buj94b}) for SiO masers. Exact methods 
	involving Accelerated Lambda Iteration have been developed for treating radiative 
	transfer in molecular systems (Jones et al. \cite{jones}; Randell et al. \cite{randell}; 
	Yates et al. \cite{yates97}) and these could be used to advantage here 
	(see Sect.~\ref{s:concs}).  However, they would be inconsistent with
	the symmetry-breaking assumption  used to simulate clumping in the present work.

\noindent (iii) a lack of any independent chemical model to calculate the SiO abundance. 
	The ratio of SiO is fixed at 10$^{-4}$ of the total particle number density, 
	where the number density is represented in terms of H$_{2}$ molecules. This
	value was chosen on the basis of the estimation given by Doel et al. (\cite{doel95}), 
        in which compositions for circumstellar gas and dust were assumed and then a gas to
	dust mass ratio argument was used to yield an upper limit to the SiO abundance.
	We note that this value is somewhat  higher than the generally accepted value
	of $\sim$5 10$^{-5}$ for the inner CE, see for example Bujarrabal et al. 
	(\cite{burra89}). 
	The effect of different SiO abundances upon SiO maser emission was investigated by
        Doel et al. (\cite{doel95}). 
	In the range 5 10$^{-5}$ -- 2 10$^{-4}$, the unsaturated
	maser gain coefficients were found to be essentially proportional to n(SiO).
        In the 
	absence of any chemical model, this abundance remains unchanged throughout these 
	calculations for all conditions encountered in the SiO maser zones for all phases 
	of the star, excepting when kinetic temperatures exceed 5700 K. On the basis
	of thermodynamic calculations, which indicate that SiO will be largely dissociated
	above this temperature in the inner CE (Field, private communication), we include 
	a crude cut-off assuming
	negligible abundance of SiO at those sites for which T$_{k}$ $>$ 5700 K.

\subsection{Maser propagation in the circumstellar envelope}
\label{ss:maserprop}

	There is good evidence that bright SiO maser spots are strongly 
	saturated in M-Miras, as discussed in detail in D95. Populations are 
	therefore substantially modified in the volume of gas occupied by bright 
	masers. Observations indicate that masers fill only a very small proportion 
	of the total volume of the maser zone (e.g. Greenhill et al. \cite{greenhill95}) 
	and the assumption is made here, as in H96, that saturating masers are 
	sufficiently spatially confined that they do not affect the general 
	pumping cycle elsewhere in the maser zone. Hence the omission of masers 
	from the calculation of level populations described in Sect.~\ref{ss:levpops}.
        Maser propagation occurs through exponential growth followed by a 
	region involving saturation, if rays achieve sufficient intensity. Maser 
	polarization (Sect.~\ref{ss:longterm} and~\ref{ss:survival}; McIntosh 
	et al. \cite{mcintosh94}; Nedoluha \& Watson \cite{ned90}, \cite{ned94}) 
	is not included in our calculations. Maser saturation and coupling of 
	the maser radiation to the kinetic master equations are treated 
	with a semi-classical formalism, in which the radiation is treated 
	according to Maxwell's equations but the response of the molecular 
	ensemble is calculated using quantum mechanical density matrix theory 
	(Field \& Richardson \cite{field84}; Field \cite{field85}; 
	Field \& Gray \cite{field88}).  All effects of saturation and 
	competitive gain are included in the propagation of a maser 
	beam through the gas containing population inversions. The masing 
	zone is assumed to amplify a black-body background at the appropriate 
	wavelength at the local kinetic temperature of the maser zone. Our 
	calculations amplify this background as a 
	function of frequency within the maser line. 

	Propagation of maser rays 
	is performed by numerical integration of the set of coupled equations

\begin{equation}\rm
\label{e:propeqn}
\frac{ d I^{\nu}_{ji}}{ d z } = A_{ji} \frac { \lambda_{ji}^{2}}{8 \pi} ( \rho_{j} - \rho_{i} ) I^{\nu}_{ji}
\end{equation}                                                              

\noindent where $I^{\nu}_{ji}$ is a specific intensity function. Maser rays are 
	assumed to be contained in a vanishingly small solid angle. Thus, in the 
	standard expression relating the angle averaged intensity to specific 
	intensity, the function  is a $\delta$-function in the angles. 
	The upper and lower level populations per sublevel, $j$ and $i$  
	respectively in Eqn.~\ref{e:propeqn}, are those which 
	respond at a certain frequency, $\nu$, within the inhomogeneously 
	broadened line. Populations as a function of frequency in the 
	presence of saturating maser radiation are calculated using the expression

\begin{equation}\rm
\label{e:semiclass}
\rho_{p} = \rho_{0p} \exp \left\{ \Sigma_{r} \left\{ T_{r}^{p} ( L , S ) I_{\nu} \right\}  \right\}
\end{equation}

\noindent originally derived in Field \& Gray (\cite{field88}) and 
	used extensively elsewhere, for example in D95. In Eqn.~\ref{e:semiclass}, 
	the $r$ subscript refers to a transition and the $p$ subscript or superscript 
	to an energy level. $T$ is 
	a maser intensity independent function of $L$ and $S$, which in turn are 
 	functions of all the non-maser radiative 
	and kinetic events built into the inversion pumping scheme. 
	The possibility that saturating maser beams may intersect one another is 
	ignored. Counter-propagating beams (or `streams') are not included in our 
	calculations (Elitzur \cite{elitzur}). The length of material traversed by 
	the maser, the `gain length', is constrained to have a maximum value equal 
	to the stellar diameter of 3.4 x 10$^{11}$ m or a value such that the velocity 
	gradient in our line-of-sight causes a Doppler shift of three Doppler widths, 
	whichever is the smaller. When the velocity shift exceeds three Doppler widths, 
	amplification almost invariably becomes negligible. In integrating the coupled Eqns.~\ref{e:propeqn} 
	and~\ref{e:semiclass}, the line-of-sight is identified by calculating the velocity 
	gradient in the line-of-sight at $r'$ and using this calculated value for maser 
	propagation. The velocity gradient in the line-of-sight, $\alpha_{los}$, may be shown 
	to be related to the radial and tangential velocity gradients by

\begin{equation}\rm
\alpha_{los}= \left( dV/dr' - V/r' \right) \sin^{2} 
\theta' \cos^{2} \phi' + V/r'
\end{equation}

\noindent Propagation of SiO masers in a velocity field may readily be performed 
	since, in Eqn.~\ref{e:propeqn}, maser propagation is treated as a function 
	of frequency within the gain profile. At each numerical integration point 
	in the propagation of masers through a masing zone, the molecular velocity 
	distribution is divided into 61 bins covering 12 Doppler widths. The 
	distribution is shifted appropriately in frequency at each integration 
	step to take account of the local velocity field. Complete velocity 
	redistribution is assumed throughout the present calculations. This is 
	achieved by summing the populations of all bins at each propagation step, 
	taking account of saturation. The summed populations are then redistributed 
	among the velocity bins according to a Gaussian profile, as described in 
	Field et al. (\cite{field94}).

\section{Results of calculations}
\label{s:results}

	Data are presented in a manner as far as possible similar to that of
	observational work described in Sect.~\ref{s:observations}. The appearance 
	of synthetic SiO maser lineshapes and images at 43 and 86-GHz in the $v=1$ 
	state is discussed,  and the predicted proper motions of bright maser components 
	are described.
	Unfortunately, there is uncertainty in
	establishing the relation between the model phase of our simulated data with
	stellar optical phase. This comes about as the exact optical phase 
	at which a pulsation-driven shock front emerges from the stellar photosphere 
	(model phase 0.0) is unknown. In order to compare
	the results of our simulations with variability observations, we calibrate 
	our results against the data for TX Cam in Sect.~\ref{r:longterm}, point (v).

\subsection{Spectra of SiO masers at 43 and 86-GHz: long term variability}
\label{r:longterm}

\begin{figure*}[t]
\vspace{0.3cm}
\hspace{0.7cm}\resizebox{17cm}{!}
{\includegraphics{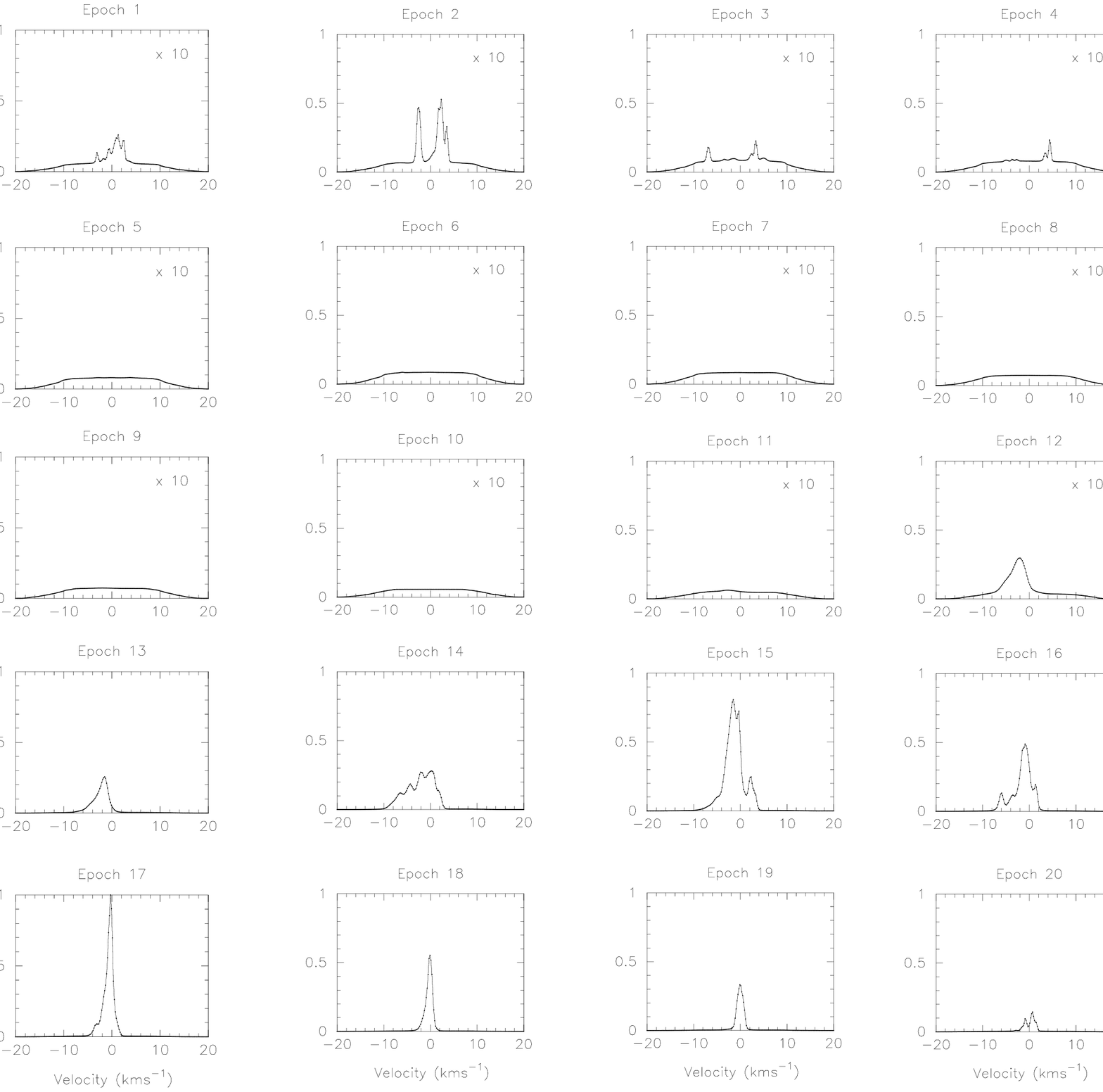}}
\vspace{-5.3cm}
\caption{\label{f:v1lines}
Time series of $v=1$ $J=1-0$ (43-GHz) SiO maser lineshapes calculated at intervals of 
16.6 days throughout the stellar cycle (of period 332.0 days).  Where ` x 10' appears, 
it indicates 
that the lineshape intensity has been multiplied up by a factor of 10, in order to show
clearly the line profile in this Figure.
The flux scale has been normalised to that of the peak lineshape value at Epoch 17.
}
\end{figure*}

\begin{figure}[t]
\vspace{0.4cm}
\hspace{0.1cm}\psfig{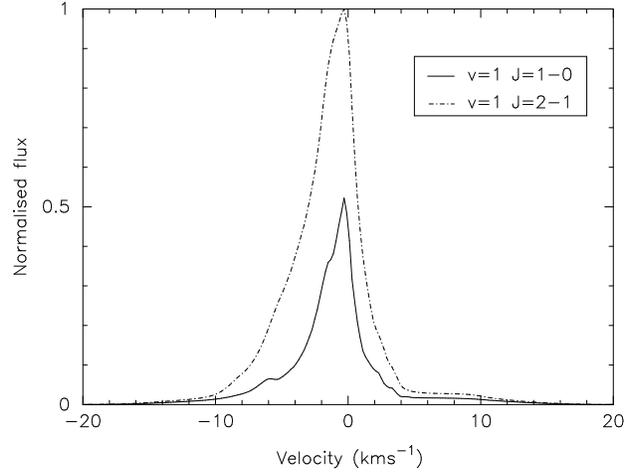}
\caption{
Cycle-averaged $v=1$ $J=1-0$ (43-GHz) and $v=1$ $J=2-1$ (86-GHz) SiO maser 
lineshapes generated from the model M-Mira.
}
\label{f:mean}
\end{figure}
\begin{figure}
\vspace{0.45cm}
\hspace{0.1cm}\psfig{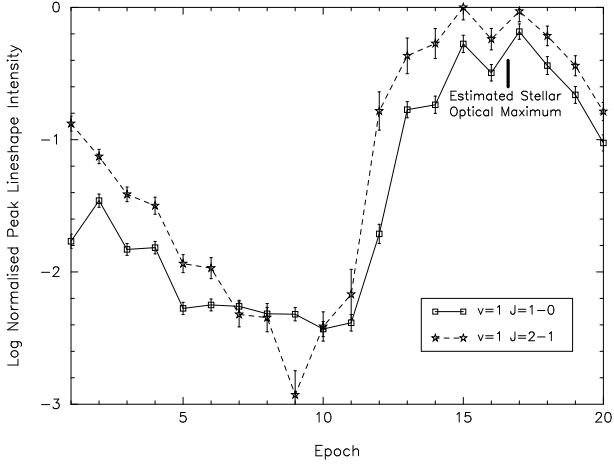}
\caption{
Peak lineshape intensity ``light curves'' for $v=1$ $J=1 - 0$ (43-GHz) and $v=1$ $J=2 - 1$ 
(86-GHz) synthetic spectra. The estimation of stellar optical maximum is
described in Sect.~\ref{r:longterm}, point (v). 
}
\label{f:newnewpeak}
\end{figure}

	We now consider the points in Sect.~\ref{ss:longterm} relating to long 
	term variability and discuss how observed and calculated features correspond. 
	Fig.~\ref{f:v1lines} shows, using the $v=1$ $J=1 - 0$ (43-GHz) maser as an example,
	how time series of maser spectra can be output from the variability simulations
	for transitions in the range $v=1$ -- $3$, $J=1-0$ -- $J=10 - 9$. These data
	may be combined to produce cycle-averaged spectra, as used by NO86, and these are
	shown in Fig.~\ref{f:mean} for $v=1$, 43 and 86-GHz lineshape calculations. 
	The variation of peak maser lineshape intensity as a function of stellar 
	phase for 43 and 86 GHz  masers is shown in Fig.~\ref{f:newnewpeak}.
	
\medskip

\noindent (i) The general forms of the data in Figs.~\ref{f:v1lines} 
	and~\ref{f:mean} are similar to those found in the most 
	extensive data available (NO86, MBA88). 

\medskip

\noindent (ii) For much  of the cycle, Fig.~\ref{f:newnewpeak} shows that the peak 
	intensity of the 86-GHz masers is greater than 
	that of the 43-GHz masers at the corresponding epoch, as typically
	observed for Miras (Pardo et al. \cite{pardo}). During epochs
	of bright maser emission  (Epochs 13 - 20), the ratio of the peak lineshape
	intensities, $v=1$ $J=2-1$/$v=1$ $J=1-0$ $\approx$ 1 - 3. Cho et al. (\cite{cho98})
	find a comparable typical value of $\approx$ 1 - 2.

\medskip

\noindent (iii) In Alcolea et al. (\cite{alcolea99}), maser minimum
	typically passes to maximum over $\sim$ 0.5 periods (i.e. the asymmetry factor
	$f$ is 0.5) for the $v=1$ 43-GHz data. Our actual values are 0.35  and 0.3 stellar periods 
	for $v=1$ 43-GHz and $v=1$ 86-GHz masers respectively. However, within the typical
	statistical uncertainties, given the number of emitting components contributing to the
	lineshape at each phase, our data is consistent with observations. In Fig.~\ref{f:newnewpeak}, 
	maser minimum may occur between Epochs 7 -- 11 and the maximum between Epochs 15 -- 17.

\medskip

\noindent (iv) The FWHM of periods of bright simulated maser intensity are 0.2 periods for 
	the 43-GHz masers and 0.25 for 86-GHz masers. These values underestimate
	observed values which lie in the approximate range 0.3 - 0.7 for the
	Miras in Alcolea et al. (\cite{alcolea99}).

\medskip

\noindent (v) By equating the model phase of maser 
	minimum light with the observed 
	optical phase of 0.67 (Sect.~\ref{ss:VLBI}) for the observations of $v=1$ 43-GHz
	masers in TX Cam,  we derive the optical maximum of our model star to correspond 
	to a model phase of 0.78, i.e. Epoch 16 - 17. Maser 
	maximum for $v=1$ $J=1-0$ emission, which occurs at Epoch 17 (phase 0.85) in our simulations,  
	therefore peaks somewhere between the optical and IR peaks if the IR peak is 
        delayed by 0.1 periods with respect to its optical counterpart. The current model is
        therefore not inconsistent with a small ($<$0.1 period) phase difference between low-frequency maser and
        IR peaks.
	Although the mean observed phase lag between optical and maser maximum light 
        is $\sim$0.1 - 0.2, we noted in Sect.~\ref{ss:longterm} (ii)
	that this value is very variable between stars and between different 
	cycles of the same star. Indeed, for 
	some cycles of some of the stars studied in Alcolea et al. (\cite{alcolea99}),
	for example R Cas and U Her, there is also negligible phase lag between stellar
	and maser maxima. 
\medskip

\noindent (vi) The contrast (ratio of maximum to minimum 
	peak lineshape intensity during the cycle) in our simulations is
	175 at 43-GHz and 850 at 86-GHz. As the observed contrasts
	are typically less than $\sim$10 $\%$ of these values, this is a clear
	failing of our simulations, though we note that very large contrasts
have been observed in $o$ Cet (NO86): see Section 2.1.(v).
In Fig.~\ref{f:v1lines} 
	it is evident  that maser emission is almost 
	eradicated between Epochs 6 to 11 (0.25 stellar periods). We 
 	discuss the reasons why the model fails in this respect in Sect.~\ref{s:concs}.

\medskip

\noindent (vii) Fig.~\ref{f:mean} shows that the bulk of emission originates
	from velocity extents of $\sim$12 kms$^{-1}$, with broader low
	intensity wings extending from around -15 to +15 kms$^{-1}$, as observed 
	by Herpin et al.\,(1998).
	Herpin et al.\,(1998) consider a number of processes which may give 
	rise to this emission, such as turbulence and asymmetric mass loss. Since 
	this CE model does not
	include turbulence and is spherically symmetric, it appears that the wings can arise  
	simply through weak emission originating from regions of the CE which are infalling
	or outflowing at relatively high velocity.

\medskip

\noindent (viii) The cycle-averaged peaks of both spectra in Fig.~\ref{f:mean} occur close to the 
	stellar velocity, peaking 0.3 kms$^{-1}$ to the blue of V$_{*}$. 
 	The cycle-averaged 
	spectral peak at $v=1$, $J=1-0$ was 0.3 kms$^{-1}$ to the red of V$_{*}$ in
	Cho et al.(\cite{cho96b}), compared 
	to the blueshift of  $\sim$1 kms$^{-1}$ found by NO86.
	Adopting the data calibration in (v), we find that in our simulated data
	$v=1$ $J=1-0$ emission is redshifted with respect to V$_{*}$ for an optical phase range
	of 0.73 -- 0.98 and blue-shifted from 0.23 -- 0.63. This does not agree with
	the statistical result of Cho et al. (\cite{cho96b}) in Sect.~\ref{s:observations} (vi).

\begin{figure*}[t]
\vspace{0.5cm}
\hspace{0.3cm}
\resizebox{16.5cm}{!}{\includegraphics{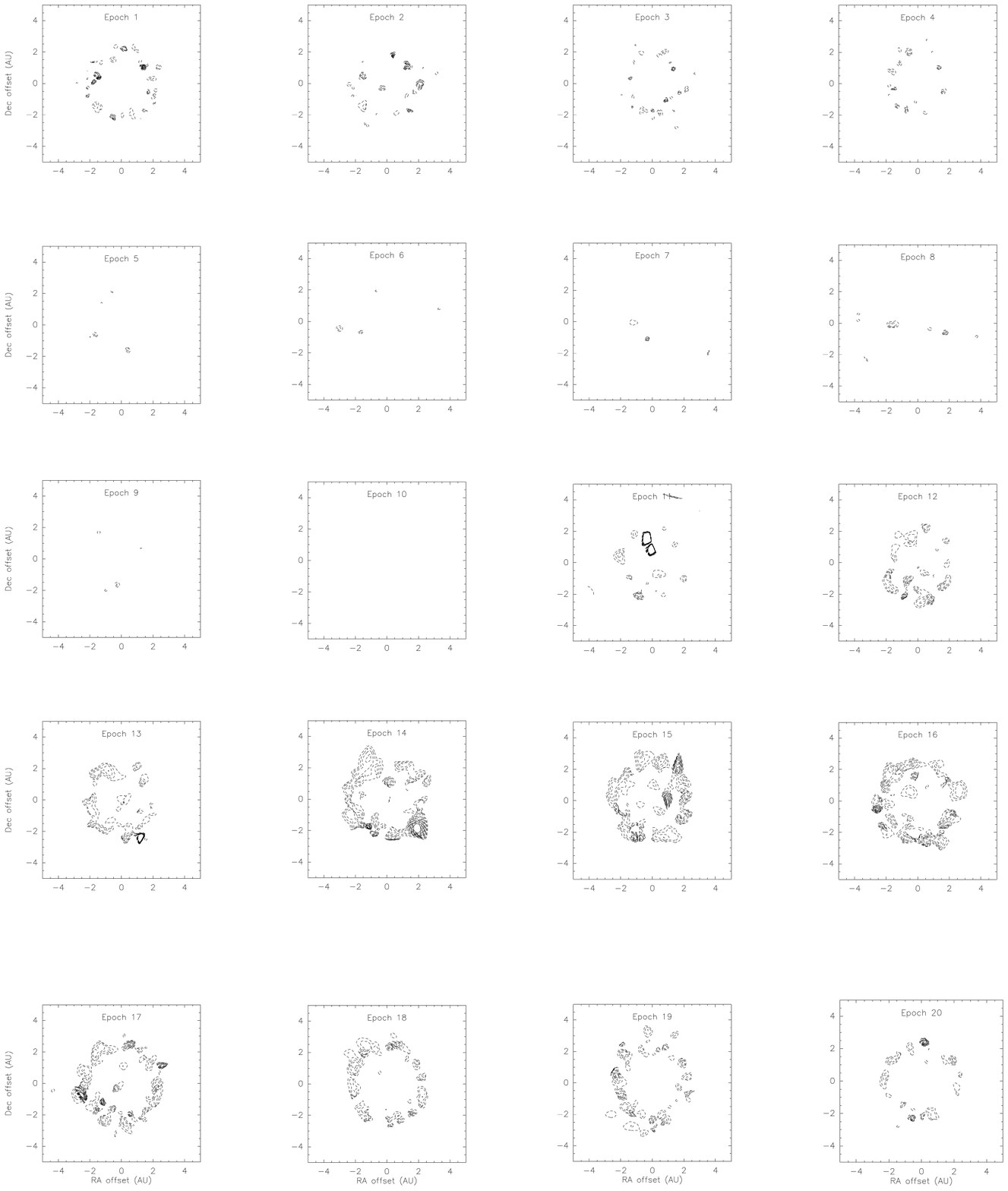}}
\caption{
	Time series of $v=1$ $J=1-0$ (43-GHz) synthetic maser images. The vertical axis 
	is the log of the velocity-integrated maser intensity, 	$I_{VEL}$. 
	Contour levels are regularly spaced over $\Delta$ log $I_{VEL}$ = 10 in order to 
	represent most fully  the broad range of component intensities in our data set.
}
\label{f:v1J1}
\end{figure*}

\subsection{Spectra of SiO masers at 43 and 86-GHz: short term variability}

	The present model  cannot deal with the very short term variability 
	of a few days or less reported in Pijpers et al.\,(\cite{pijpers94}),  
	which the authors attribute to short wavelength sound waves generated 
	through convection, a phenomenon not included in the present stellar 
	pulsation model (see Pijpers \& Hearn \cite{pijpers89}). However
	conditions in the CE do change significantly between model epochs,
	on timescales of $\sim$17 days.
	This is clearly illustrated by Fig.~\ref{f:v1lines} and~\ref{f:newnewpeak}.

\begin{figure}
\vspace{0.1cm}
\hspace{0.0cm}\psfig{file=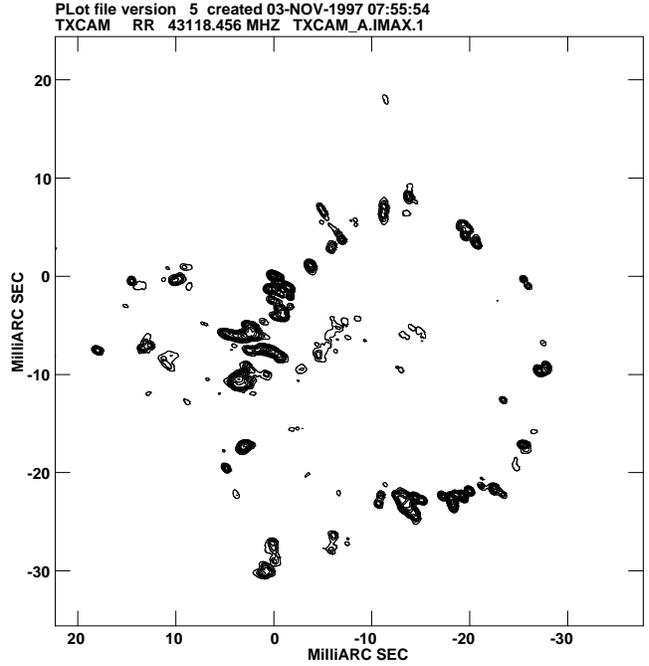,height=10.3cm,angle=0}
\vspace{-0.6cm}
\caption{
	A spectrally-averaged VLBA$^{1}$ image of the 43-GHz $v=1$ $J=1-0$ SiO masers 
	around TX Cam. The image covers a velocity range of 
	3 to 15 kms$^{-1}$. The peak in 
	the averaged beam is 19.387 Jybm$^{-1}$, the contours are plotted at levels of
	0.5 Jybm$^{-1}$ multiplied by factors of (-2.68,-1.93,-1,1,1.931,2.683,3.728,5.179,
	7.197,10,13.89,19.31,26.83,37.28,51.79,71.79,100).  The synthesised restoring beam 
	is 0.5 by 0.4 mas at a position angle of 20 degrees. $^{1}$The Very Long 
Baseline Array (VLBA) is operated by the National Radio Astronomy Observatory under cooperative agreement with the National Science Foundation (NSF).
}
\label{f:txcam_obs}
\end{figure}

\begin{figure}[t]
\vspace{0.6cm}
\psfig{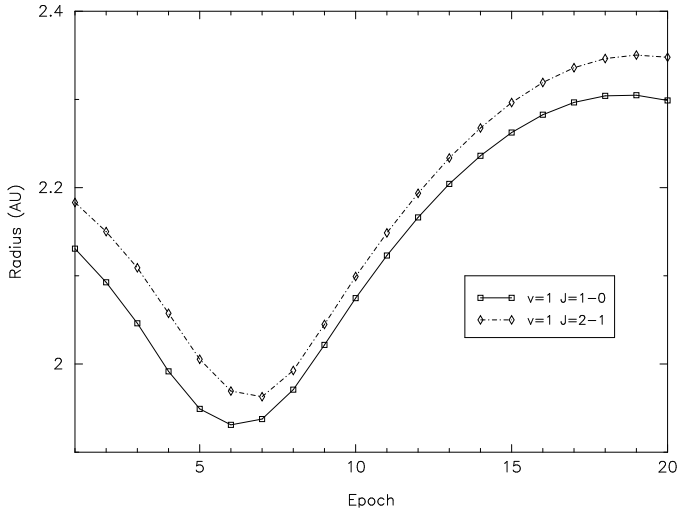}
\caption{
Mean ring radius for $v=1$ $J=1-0$ (43-GHz) and $v=1$ $J=2-1$ (86-GHz) masers, 
generated by the model M-Mira variable, as a function of stellar phase. The 
ring radius is determined by taking the average radius over the fifty brightest
components at each epoch. The uncertainty in these values is typically about 0.13 AU. 
}
\label{f:radmotions}
\end{figure}

\subsection{Synthetic images at 43-GHz and 86-GHz: survival and 
proper motion of maser components}
\label{r:images}

	H96 showed that masers at model phase zero (Epoch 1) were disposed in an 
	approximate ring around the host star at a radius of $\sim$1 R$_{*}$ 
	from the photosphere. 
	The present work shows that this ring expands, contracts, undergoes severe disruption  and 
	reappears according to the stellar phase. 
	Fig.~\ref{f:v1J1} shows the time series of synthetic images resulting 
	from our simulations at 43-GHz, which can be displayed for any transition. 
	The interval between
	epochs in Fig.~\ref{f:v1J1} is 16.6 days, corresponding
	to 0.05 stellar periods. The choice of 1500 maser sites in order to generate 
	these images was based on numerical experiments at Epoch 1. It would now appear 
	that a smaller number might have been more appropriate, since the ring 
	structures in Epochs 14 -- 18 are rather more complete than observations 
	(of TX Cam) currently suggest.
	Fig.~\ref{f:txcam_obs} shows a VLBA image
	of the 43-GHz masers in TX Cam, for comparison with our simulated data.
	Fig.~\ref{f:radmotions} shows maser ring radii for 43 and 86-GHz masers
	as a function of the stellar pulsation phase.
	 At all events a number of useful conclusions 
	can be drawn from these data.

	\medskip

	\noindent (i)  In Fig.~\ref{f:v1J1}, it is evident that tangential amplification is the norm, 
	but that an occasional feature may form over the disk of the star, as at
	Epoch 17 at 43-GHz. Data for TX Cam indeed show that an occasional maser 
	spot may form over the disk of the star, as shown in Fig.~\ref{f:txcam_obs}.

	\medskip 

	\noindent (ii) The radius of the 43-GHz "ring" of masers 
	varies between 1.93 -- 2.3 AU in Fig.~\ref{f:radmotions}. The average radius of the maser ring declines 
	between Epochs 1 -- 6, at a mean rate of 0.2 AU in 83 days, 
	corresponding to 4.2 kms$^{-1}$ (cf. infall of 4 kms$^{-1}$ in R Aqr over a similar
	duration of the stellar cycle).
	Contraction of the ring is followed by a larger shock-driven expansion at a
	rate of 3.2 kms$^{-1}$ (cf. dominant expansion in the TX Cam observations, at a 
	typical velocity of  3.65 kms$^{-1}$). In our simulations, 
	expansion of the ring is accompanied by a period of bright maser emission, induced
	by the temperature and density enhancements of the post-shock gas.

	\medskip

	\noindent (iii) The radius of the $v=1$ 86-GHz ring of masers is consistently similar, 
	to within 0.1 AU, to that of the $v=1$ 43-GHz ring throughout the stellar cycle. 
	Gray \& Humphreys (\cite{gray2000}) show that this is not the case for masers
	in different vibrational states. The radial motions of the rings
	are highly coupled as bright $v=1$, 43 and 86-GHz maser emission often originates from 
	shared components in the CE. This is supported by observational data of Colomer 
	et al.(\cite{colomer96}) and Doeleman et al.(\cite{doele}), although their
	multi-transition results were not made simultaneously.

	\medskip

	\noindent (iv) More 
	sites in the CE produce
	maser emission at 43-GHz (16 $\%$ of the 1500 sites at Epoch 15) than at  
	86-GHz (7.5 $\%$ at the same epoch). However, of the components which do emit, 
	$v=1$ $J=2-1$ emission is typically
	significantly stronger than that of at $v=1$ $J=1-0$, leading to the more
	intense images and lineshapes in our data. (There is a general
	trend in our data for fewer emitting components as a function of increasing
	$J$-value for the transition.) It is not clear why the $v=1$ $J=2-1$ should be the stronger. 
	Results in D95 also suggested that 43 and 86-GHz emission in $v=1$ form under similar 
	conditions. As discussed in D95, competition betwen masers and cycling of 
	populations between the $v=1$ and $2$ and $J=1$ and $2$ states result in a complicated 
	interplay whose outcome determines whether the 43 or 86-GHz line will emerge as 
	a strong maser.

	\medskip 

	\noindent (v) In our simulated data, the passage of a shock front
	through the maser zone causes minimum maser light. The combination of
	physical conditions in the zone are unsuitable for yielding
	bright maser emission, in particular, the high temperatures which dissociate
	SiO. Although temperatures in real stars are unlikely to be as high as in this model,
	we believe that the general picture provided here is the correct one.
	That is to say a shock front disrupts the existing maser ring and
	new features then form in the gas in the wake of the shock i.e. a new maser ring
	will appear to form which has a smaller angular extent than the previous
	ring.



\section{Concluding remarks}
\label{s:concs}

	These simulations have some rather severe limitations, which we discuss
	below. However, a significant conclusion of the present work is that
	coupling a SiO maser model to a pulsating AGB stellar model, keeping
        the stellar IR radiation field constant throughout the cycle,   
	reproduces qualitatively much  of the available observational data.
	This is particularly evident in the location, tangential amplification  and proper motion
	of SiO masers (although clearly our spherically symmetric model
	will not reproduce asymmetrical effects) in the CE.  The 
	essential point is that shocks in the inner CE have an effect on
        SiO maser emission which, given the lack of time-dependent chemistry
	and efficient molecular cooling in our present simulations, could explain much of the
        SiO maser variability phenomenon. Of course the inclusion of a varying
	stellar IR radiation field, in addition to the improvements outlined above,
	is essential for modelling fully the environments provided by such stars.
	We note that it is not yet established whether SiO masers are predominantly
	radiatively or collisionally pumped.

	A  number of important quantitative shortcomings are evident in our simulations, 
	which may be summarised as follows.  The duration of the period 
	of bright maser emission is underestimated, the contrast in the maser lightcurve is
	overestimated, maser emission is very weak for a significant portion of
	the stellar cycle and in observed data, red-shifted emission dominates for the
	greater portion of the stellar cycle, whereas in the simulated data blue-shifted
	emission dominates.
	In this discussion 
	we should recall that the model parameters set out in Table~\ref{t:mirainputs} 
	are appropriate to $o$-Ceti, and these represent an arbitrary choice which may 
	not necessarily yield the behaviour which may be characterised as ``typical'' of 
	M-Miras. 

        An important problem remains the difficulty in relating accurately 
	the phase of our model star to stellar optical phase. As we have stated, the current
        hydrodynamic model is not really capable of deciding this issue because it has a 
        fixed-temperature photosphere. VLBI observations, however, are at present consistent
        with the view that the shock wave in the envelope arrives at the maser zone in a
        reasonably well-constrained phase range for a group of objects with very different
        stellar parameters. Assuming that the observed maser ring in a VLBI experiment corresponds
        well to the radius of the shock at the time the shock impacts the maser zone - as is
        the case in our model - we deduce from the five stars in Table~2, that the mean
        phase, $\bar{\phi}$ of shock arrival is 0.71 in the modellers' definition, with a standard         deviation
        on the
        mean, $(\sum_{j=1}^{n} (\phi_{j} - \bar{\phi})^{2}/(n(n-1)))^{1/2}$, equal to only 0.029
	periods.
	In the case of R Cas, Phillips et al. (\cite{phillips2001})
         give sufficient data to make a reasonable estimate of the root-mean-square 
         uncertainty in the time required for the shock to reach the maser
         zone. The error in the angular
         measure is dominated by the uncertainty in the stellar diameter, quoted as
         30$\pm$6 mas. The fractional error in the angular distance to be crossed 
         by a shock travelling to the maser ring, of diameter 57 mas,  is
         therefore 6/27 = 0.22. To this we add the fractional error in the distance
         to R Cas, about 0.2 from the authors' figures, to get 0.44. This level of 
         uncertainty is certainly large, 
         but we are basing our results on a group of five, not on a single source. 
         Assuming that 0.44 is a reasonable fractional error for all five objects, 
	we obtain a root-mean-square uncertainty on the mean for the five objects 
         in  Table~2 of 0.15 periods. This is certainly larger than the standard 
         deviation on the mean, but is still usefully restrictive. 
        This information is quite independent of any assumed pumping scheme, as it is based
        only on observation.
        Given that most objects have a roughly fixed delay from shock impact to maser maximum, and
        a further constant phase shift links the modellers' and optical phase definitions
        (see Section 4.1.5) then we expect a fairly constant phase shift between the peaks
        of the optical
        (or IR) light curves and the maser peak. Observationally, this shift is close to 0.0
        in the IR case for most, but not all objects. This link would be quite natural if the
        shock and IR photosphere are dynamically linked in many objects. Such a link is plausible
        because much of the opacity in the 3-10 $\mu$m range is derived from water molecules.
        The shock would provide a dense layer which could be optically thick for part of the
        stellar cycle. Observational support for this view comes from observations by 
        Yamamura, de Jong \& Cami (\cite{yama}) who found that the warmer of two water layers, located
        at around two nominal stellar radii in $o$ Cet was optically thick. The high temperature
        suggests the layer is shock-heated, and the mid-IR photosphere would therefore lie
        somewhere in the post-shock gas. 

\begin{table}

\begin{center}
\vspace{0.5cm}
 \begin{tabular}
 {lllll}
\hline
                               &      \\
\multicolumn{5}{c}{Approximate Phases of Shock Arrival}\\
                               &      \\
 \hline
\\
Object                         &  R$_{ring}$  (AU) & R$_{*}$ (AU)     &   P (d)       &  Phase\\
\hline
\\
U Her$^{1}$                    &      4.3          &      2.4         &    455        &  0.65 \\
TX Cam$^{2}$                   &      4.9          &      1.8         &    557        &  0.77 \\
IRC+10011$^{3}$                &      5.5          &      1.8         &    650        &  0.79 \\
R Aqr$^{4}$                    &      3.0          &      1.14        &    387        &  0.66 \\
R Cas$^{5}$                    &      4.56         &      2.4         &    430        &  0.69 \\
\\
\hline
\end{tabular}
\end{center}
\caption{\label{t:shockphase}
Shock arrival data for Miras observed at VLBI resolution: columns are the object name, the radius
of the observed maser ring, an estimate of the stellar radius, the stellar period, and the
approximate phase of arrival of the shock, assuming the period-averaged speed from our model
of 0.00723\,AU per day. Notes: $^{1}$, data from Diamond et al. \cite{diamond94}; $^{2}$, from
Desmurs et al. \cite{desmurs} with stellar radius from Diamond et al. \cite{diamond94}; $^{3}$, from
Desmurs et al. \cite{desmurs} with radius assumed equal to that of TX Cam; $^{4}$, from Hollis et al.
\cite{hollis} with radius assumed equal to the model star (see Table~1); $^{5}$, data from
Phillips et al. \cite{phillips2001}.
}
\end{table}


	A significant source of error in these calculations is
	likely to be the lack of molecular coolants such as CO in this 
	model.
        The shocks in the present work (see Fig.~\ref{f:bowenplot}a) are quite close 
	to being adiabatic. The effect of coolants, such as CO is not included 
	and cooling of the shocks in the present model takes place broadly on the hydrodynamic 
	timescale. This timescale depends on the phase of the star and lies in the approximate 
	range of 60-100 days. If the effects of cooling by CO are included (Cuntz \&  
	Muchmore \cite{cuntz}; Woitke et al. \cite{woitke}), the shocks cool on a radiative timescale 
	of only 1 -- 2 days and the shocks are much closer to isothermality than those 
	used in the present work. This means that SiO maser emission would not
	be severely weakened for a large portion of the stellar cycle, as it is
        in the current work, partly due to the dissociation of SiO. We would therefore expect more
        accurate shock conditions, and possibly time-varying IR radiation, to reduce the maser
        dynamic range from the high levels found in the simulations.
        Time-dependent chemistry 
	is also a requirement for providing the abundance of SiO and of other coolant species.

        A number of the discrepancies mentioned above may also arise from the use of a spherical 
	pulsation model. This yields too clean-cut behaviour of the CE. Variations in the 
	physical conditions in the CE almost certainly do not take place in the same manner 
	equally in all directions. Moreover clumping of material (Sect.~\ref{r:longterm}) 
	due to thermal 
	instabilities is inherently spatially inhomogeneous.  Given non-spherical behaviour 
	(Sect.~\ref{ss:VLBI}), the CE at any one phase of stellar pulsation will 
	exhibit a spread of physical conditions for any given radius, rather than the unique 
	set used here. This will increase the duration of bright emission and reduce the contrast 
	between maximum and minimum maser brightness. In this connection the VLBA data on TX Cam 
	at a variety of phases show that the entire maser ring never fades away in its entirety 
	as our present models suggest (Diamond: private communication). The rings wax and wane in 
	intensity but $\sim$50\% of the ring generally remains visible. An example of a prediction 
	of the model that is verified by observations of maps at different stellar phases is 
	that maser emission is not exclusively tangential and that an occasional maser spot 
	may indeed be found over the disk of the host star, as we predict.

        To improve our understanding of the extended atmospheres of AGB stars, the following are 
	necessary. The relationship between the optical and model phases should be reliably fixed 
	by a combination of interferometry observations and theory.  The hydrodynamic pulsation 
	model should in future incorporate a varying IR radiation field, which in turn requires
        the hydrodynamic model to incorporate an accurate prediction of the radius of the IR
        photosphere as any phase. Additionally, we require
        some chemical modelling of SiO formation, of the H, 
	H$_{2}$ 
	balance and of coolant molecules such as CO and H$_{2}$O. As noted above, this is 
	significant in determining the radiative timescale for cooling of the shocks in the CE. 
	Chemical modeling also has implications both for the relative amplification factors at 
	different phases and for the nature of the collisional pumping mechanism, since H and 
	H$_{2}$ have significantly different rate coefficients for rotationally and 
	rovibrationally inelastic collisions with SiO in the new calculations by 
	Jimeno et al. (\cite{jim}). In addition, 
	a time-dependent description of silicate, that is, olivine dust formation 
	should be introduced, noting that dust formation also affects the local thermal 
	balance. At present, the importance of including magnetic field effects in the CE is unclear.
	With respect to the maser model, 
	accelerated lambda iteration methods should be used to replace LVG 
	(Jones et al. \cite{jones}; Randell et al. \cite{randell}; Yates et 
	al. \cite{yates97}). 
	Accelerated lambda iteration methods 
	have the important characteristic that they can incorporate the velocity, 
	temperature and number density structure of the CE explicitly into the calculation 
	of the populations of SiO rovibrational levels. These methods require development 
	for a spherical geometry for molecular systems and are computationally expensive, 
	though not prohibitively. New rate coefficients for collisional energy transfer between 
	H$_{2}$ and SiO must also be computed. The outstanding observational 
	requirement is for more sets of VLBI images disposed at equal intervals of $\sim$0.1 
	in phase (near-)simultaneously for several SiO maser transitions towards suitable 
	 Mira long-period variables.

\begin{acknowledgements}
EMLH thanks the Swedish Foundation for International Co-operation
in Science (STINT) for the financial support of this project.
\end{acknowledgements}

\end{document}